\documentclass{article}
\usepackage{spconf,amsmath,graphicx,amsfonts}

\title{Toward Efficient Hyperspectral Image Processing inside Camera Pixels}
\name{Gourav Datta$^1$, Zihan Yin$^1$, Ajey Jacob$^2$, Akhilesh R. Jaiswal$^{1,2}$, Peter A. Beerel$^1$}
\address{$^{1}$University of Southern California, $^{2}$Information Sciences Institute} 
\begin{document}
%
\maketitle
\begin{abstract}
Hyperspectral cameras generate a large amount of data due to the presence of hundreds of spectral bands as opposed to only three channels (red, green, and blue) in traditional cameras. This requires a significant amount of data transmission between the hyperspectral image sensor and a processor used to classify/detect/track the images, frame by frame, expending high energy and causing bandwidth and security bottlenecks. To mitigate this problem, we propose a form of processing-in-pixel (PIP) that leverages advanced CMOS technologies to enable the pixel array to perform a wide range of complex operations required by the modern convolutional neural networks (CNN) for hyperspectral image recognition (HSI). Consequently, our PIP-optimized custom CNN layers effectively compress the input data, significantly reducing the bandwidth required to transmit the data downstream to the HSI processing unit. This reduces the average energy consumption associated with pixel array of cameras and the CNN processing unit by $25.06\times$ and $3.90\times$ respectively, compared to existing hardware implementations. 
Our custom models yield average test accuracies within $0.56\%$ of the baseline models for the standard HSI benchmarks.


\end{abstract}

\section{Introduction}
\label{sec:intro}

3D image recognition has been gaining momentum in the recent past \cite{alhamzi2014survey}, with applications ranging from augmented reality \cite{lv2021augmented} to satellite imagery \cite{facciolo2017satellite}. In particular, hyperspectral imaging (HSI), which extracts rich spatio-spectral information about the geology at different wavelengths, has shown vast promise in remote sensing \cite{chen2014survey}, and thus, has become a prominent application for 3D image classification. In hyperspectral images (HSIs), each pixel is typically denoted by a high-dimensional vector where each entry corresponds to the spectral reflectivity of a particular wavelength \cite{chen2014survey}, and constitutes the $3^{rd}$ dimension of the image. The aim of the recognition task is to allocate a unique label to each pixel \cite{zheng2020fpga}. 

\begin{figure}[!t]
\centering
\includegraphics[width = 0.98\linewidth]{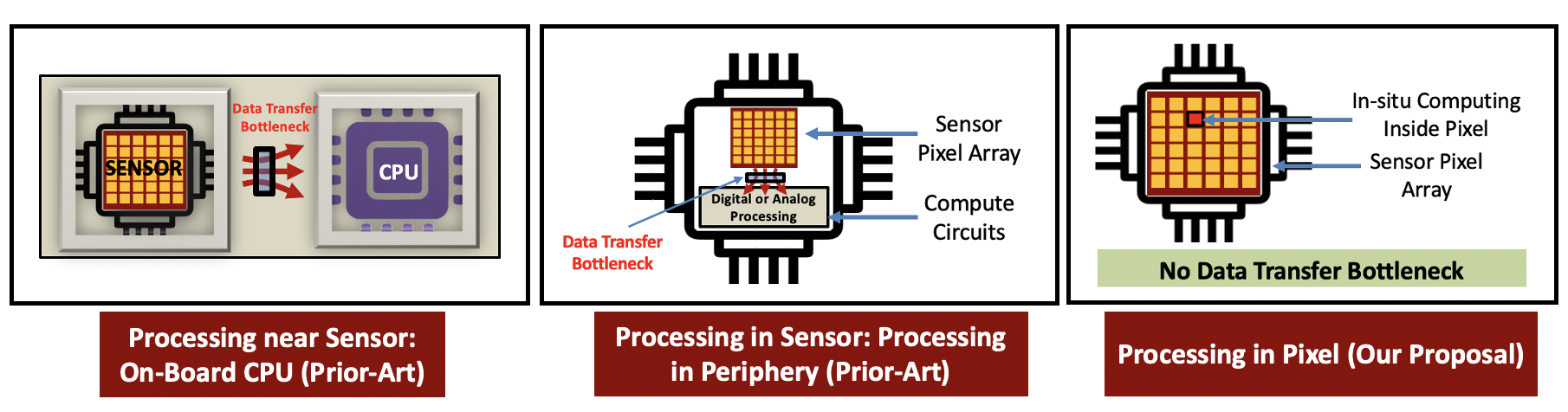}
\caption{Existing (prior art) and proposed solutions to alleviate the energy, throughput, and bandwidth bottleneck in HSI caused by the segregation of \textit{sensing} and \textit{compute}.}
\label{fig:pip_initials}
\vspace{-3mm}
\end{figure}

In comparison to 2D CNNs that are used for classifying traditional RGB images \cite{krizhevsky2012alexnet}, multi-layer 3D CNNs require significantly higher power and storage costs \cite{li2016cost}. Additionally, the high resolution and spectral depth of HSIs necessitates a large amount of data transfer between the image sensor and the accelerator for CNN processing, which leads to energy and bandwidth bottlenecks. 
To alleviate these concerns, our proposal first described in \cite{datta2022scireports} for 2D TinyML workloads and illustrated in Fig. \ref{fig:pip_initials}, presents a pathway to embed complex computations, such as convolutions and non-linear activation functions, inside and in the periphery of the pixel array respectively, using algorithm-hardware co-design catered to the needs of modern HSI applications.

Due to the size and compute limitations of current in-sensor processing solutions, it is not feasible to execute state-of-the-art (SOTA) 3D CNN models for HSI completely inside the pixel array. Fortunately, the SOTA HSI models \cite{roy2020hybrid,zhe2021hsi} are not as deep as modern CNN backbones used for traditional vision tasks, such as ResNet-$50$ \cite{he2015deep}, which increases the potential benefit of optimizing the first few layers using PIP. This motivates studying how the image sensor can effectively compress the HSIs by implementing the first few 3D CNN layers using PIP. \textit{Towards this goal, we present two CNN models for HSI that captures our PIP hardware and achieve significant compression obtained via a thorough exploration of the 3D CNN algorithmic design space.}  
Our models yield a reduction of data rates 
(and corresponding power consumption) between the in-pixel front-end and back-end CNN processing by up to $10.0\times$. 

The remainder of the paper is organized as follows. Section \ref{sec:prior_work} discusses existing works on energy-efficient in- and near-sensor processing approaches and accurate CNN models for HSI. Section \ref{sec:algo_HW} presents our proposed algorithm-hardware co-design for PIP-implemented HSI models. Section \ref{sec:results} then discusses the test accuracy and energy-efficiency of our proposed models. Finally, we provide a summary of our contributions in Section \ref{sec:conc}.

\section{Related Work}
\label{sec:prior_work}

\subsection{Energy-efficient on-device vision}

To address the compute-efficiency, latency, and throughput bottlenecks of 2D computer vision algorithms, recent research have proposed several processing-near-sensor (PNS) \cite{kodukula2020sensors, sony2020vision},  
processing-in-sensor (PIS) \cite{chen2020pns}, and PIP solutions \cite{scamp2020eccv, Mennel2020UltrafastMV}. 
%
PNS approaches incorporate a dedicated machine learning (ML) accelerator chip on the same printed circuit board \cite{kodukula2020sensors}, or 2.5D/3D stacked with a pixel chip \cite{sony2020vision}. PIS approaches, in contrast, leverage parallel analog computing in the peripheral circuits of a memory array \cite{chen2020pns}. 
However, existing PIS solutions require serially accessing the CNN filters and PNS solutions demand high inter-chip communication, both of which incur significant 
energy and throughput bottlenecks. 
Moreover, some existing PIP solutions, such as SCAMP-5 \cite{scamp2020eccv} consist of CMOS-based pixel-parallel SIMD processor arrays that offer improved throughput but are limited to implementing simple networks consisting of  fully-connected layers.  Other PIP solutions, based on beyond-CMOS technologies \cite{Mennel2020UltrafastMV}, are highly area-inefficient, and currently not compatible with existing CMOS image sensor manufacturing processes, which limits support for multiple channels, confining their application to simple ML tasks 
such as character recognition. 
Thus, all these existing approaches have significant limitations when applied to deep 3D CNN models for HSI. 


\subsection{CNN models for HSI}

The 2D CNN autoencoder, proposed in \cite{chen2014survey}, was the first attempt to extract deep learning features from the compressed latent space of HSIs. 
Subsequently, researchers have proposed feeding each target pixel data along with the data associated with neighboring pixels (a patch of size $n{\times}n$ with the input pixel at the center, where $n$ is odd) and creating a 3D CNN architecture that better captures the dependencies between the spectral and spatial features \cite{hamida2018-3d} and achieves improved accuracy compared to earlier efforts. 
More recently, \cite{roy2020hybrid,luo2018hsi} successfully fused these 3D extracted features using several 2D convolutional layers to extract a more fine-grained representation of spectro–spatial information and achieve even higher accuracy for some HSI benchmarks. 

For the purpose of this paper, we would like to emphasize that, compared to conventional 2D CNNs, the first few layers of these HSI 3D models are memory- and compute-intensive due to the existence of the multiple spectral bands and the need to stride the filter across the third dimension. 


\begin{figure}
\centering
\includegraphics[width = 0.98\linewidth]{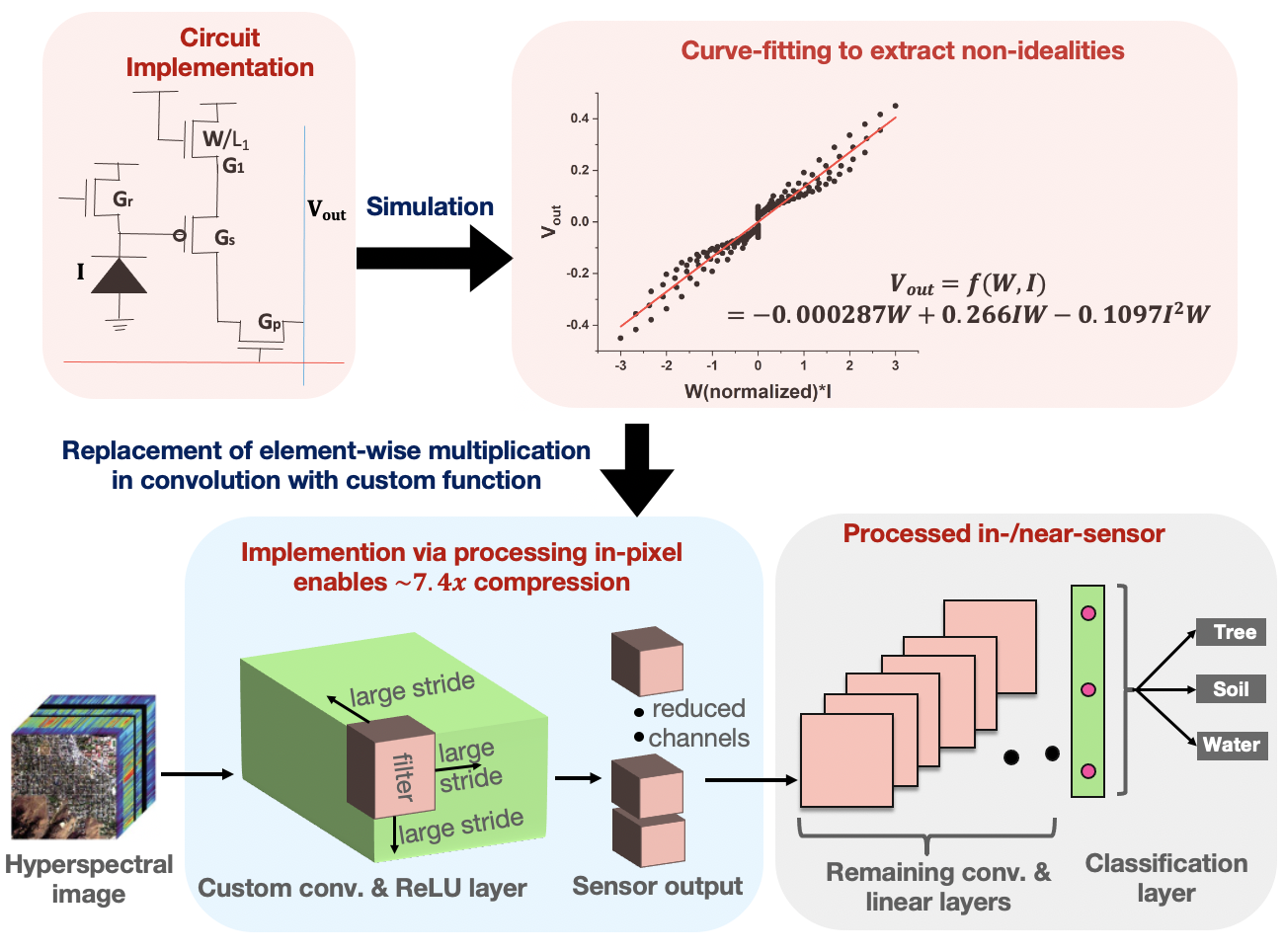}
\caption{Algorithm-hardware co-design framework that enables our proposed PIP approach to optimize both the performance and energy-efficiency of HSI.}
\label{fig:pip_framework}
\vspace{-4mm}
\end{figure}

\section{Algorithm-Hardware Co-Design for HSI}
\label{sec:algo_HW}

\subsection{Preliminaries}

We propose to implement the first 3D convolutional layer of a HSI model by embedding appropriate weights inside pixels  which includes the spatial, spectral and output channel dimensions as shown in Fig. \ref{fig:pip_framework}. The weights are encoded as transistor widths which are fixed during manufacturing\footnote{The weights can also be programmable by mapping to emerging resistive non-volatile memory elements embedded within individual pixels \cite{datta2022scireports}.}. However, this lack of programmability is not necessarily a problem as the first few layers of a HSI model extract high level spectral features that can be common across various benchmarks.
By activating multiple pixels simultaneously, the weight modulated outputs of different pixels are summed together in parallel in the analog domain, effectively performing a convolution operation. Our approach also leverages the existing on-chip \textit{correlated double sampling} circuit of traditional image sensors to accumulate both negative and positive weights that are required to train accurate HSI models and implement the subsequent ReLU operation. The circuit details of our PIP implementation can be found in \cite{datta2022scireports}. 
\vspace{-2mm}
\subsection{PIP-optimized HSI models}

The number of convolutional output channels, range of weights, and bit-width for ReLU output impact the models' accuracy. However, these parameters are tightly intertwined with PIP specific circuit implementations.  Moreover, our CNN model needs to capture the non-idealities of our PIP hardware that exist because the conductance of the transistors in the pixel array, which implement the first layer weights, do not vary perfectly linearly with pixel output. 
We thus propose a tightly intertwined algorithm-hardware co-design framework that is illustrated in Fig. \ref{fig:pip_framework}. In particular, we simulate the pixel output voltage with varying weights and inputs, the latter reflecting the photo-diode currents. We use standard curve-fitting tools to obtain the custom element-wise function shown in Fig. \ref{fig:pip_framework}. We then create a custom CNN layer based on our circuit behavior mapped curve-fitting models. This custom layer replaces the normal convolutional layer in our algorithmic framework, thereby capturing the accurate circuit behavior. 


\subsection{PIP-enabled compression}\label{subsec:pip_compression}

In order to compress HSIs and improve the area-efficiency of our PIP implementation, we propose to limit our first convolutional layer to have a reduced number of output channels and large strides. Fortunately, we show in Section \ref{sec:results} that this does not lead to a significant drop in accuracy compared to existing SOTA HSI models. Moreover, we propose to constrain the first ReLU layer to be quantized with relatively low precision to maximize compression and reflect the energy-accuracy tradeoff due to limited precision on-chip ADCs. To achieve this goal, we use the popular quantization-aware training (QAT) method \cite{courbariaux2016binarized, datta2021hsi}. 
Here, the ReLU output tensor is fake-quantized in the forward path \cite{courbariaux2016binarized}, while the gradients of the convolutional and ReLU layers are computed using straight through estimator \cite{courbariaux2016binarized} which approximates the derivative to be equal to $1$ for the whole output range.

To quantify the compression ($C$) obtained by the first 3D convolutional and ReLU layer implemented by PIP, assume $X{\in} \mathbb{R}^{H^i{\times}{W^i}{\times}{C^i}{\times}{D^i}}$ is the input hyperspectral image, $O^l{\in} \mathbb{R}^{H^o{\times}{W^o}{\times}{C^o}{\times}{D^o}}$ is the ReLU output, and $N$ is the bit-precision of the ReLU activation map obtained by our training framework. $C$ can then be computed as
\begin{equation}\label{eq:pip_compression}
    C = \left(\frac{{H^o{\times}{W^o}{\times}{C^o}{\times}{D^o}}}{{H^i{\times}{W^i}{\times}{C^i}{\times}{D^i}}}\right)\cdot\frac{12}{N}
\end{equation}
\noexpand
where $H^i$, $W^i$, $C^i$, $D^i$ denote the height, width, \#  of channels (typically equal to $1$ for HSIs), and \#  of spectral bands of the image, respectively. Note that the factor $\frac{12}{N}$ arises because the traditional camera pixels have a bit-depth of $12$ \cite{onsemi:AR0135AT}.

Similarly, $H^o$, $W^o$, $C^o$, $D^o$ represent these dimensions for the ReLU output. While $C^o$ is obtained directly from our training framework discussed above, $H^o$, $W^o$, and $D^o$ are computed as follows
\begin{equation}\label{eq:input_output_mapping}
    Z^o = \left(\frac{Z^i-k+2p}{s_{Z}} +1\right)
\end{equation}
where $Z$ represents the height ($H$), width ($W$) or spectral band ($D$). Here, $k$ represents the filter size, $p$ represents the padding, and $(s_{H}, s_{W}, s_{D})$ represents the stride dimensions. 

\begin{table}
\begin{center}
\scriptsize\addtolength{\tabcolsep}{-1.2pt}
\begin{tabular}{|c|c|c|c|c|c|}
\hline
Dataset & Architecture & $C_l^o$ & $(s_H, s_W, s_D)$ & $N$ & $C$ \\
\hline
Indian Pines & CNN-3D & 2 & (1, 1, 3) & 6 & 8.33  \\
\hline
Salinas Scene & CNN-3D & 2 & (1, 1, 3) & 8 & 6.25   \\
\hline
HyRANK & CNN-3D & 2 & (1, 1, 3) & 5 & 10.00  \\
 & CNN-32H & 4 & (1, 1, 3) & 5 & 5.00 \\
\hline
\end{tabular}
\end{center}
\vspace{-3mm}
\caption{Values of the training hyperparameters that lead to maximum compression in our custom HSI models over the three datasets.} 
\label{tab:hsi_compression}
\vspace{-3mm}
\end{table}

\begin{table}
\begin{center}
\scriptsize\addtolength{\tabcolsep}{-1.2pt}
\begin{tabular}{|c|c|c|c|c|}
\hline
Authors & Architecture & AA ($\%$) & OA ($\%$)
 & Kappa ($\%$) \\
\hline
\hline
 \multicolumn{5}{|c|}{Dataset : Indian Pines} \\
\hline
Song et al. (2020) \cite{song2018feature} & DFFN & 97.69 & 98.52 & 98.32   \\
\hline
Zhong et al. (2018) \cite{zhong2018hsi} & SSRN & \textbf{98.93} & \textbf{99.19} & \textbf{99.07} \\ 
\hline
This work & B-CNN-3D & 98.66 & 98.54 & 98.30  \\
& C-CNN-3D & 98.12 & 98.08 & 97.77 \\
\hline
\hline
 \multicolumn{5}{|c|}{Dataset : Salinas Scene} \\
\hline
\hline
Song et al. (2020) \cite{song2018feature} & DFFN & 98.75 & 98.87 & 98.63   \\
\hline
Meng et al. (2021) \cite{zhe2021hsi} & DRIN & 98.6 & 
96.7 & 96.3 \\
\hline
This work & B-CNN-3D & \textbf{99.30} & \textbf{99.28} & \textbf{99.18} \\
& C-CNN-3D & 98.82 & 98.75 & 98.57 \\
\hline
\hline
 \multicolumn{5}{|c|}{Dataset : HyRANK} \\
\hline
\hline
Meng et al. (2021) \cite{zhe2021hsi} & DRIN & 56.0 & 54.4 & 43.3 \\
\hline
This work & B-CNN-3D & 60.88 & 52.10 & 44.68 \\
& C-CNN-3D & 60.93 & 51.18 & 47.90 \\
\hline
This work & B-CNN-32H & \textbf{64.11} & \textbf{63.30} & 52.94 \\
& C-CNN-32H & 63.73 & 62.36 & \textbf{53.17} \\
\hline
\end{tabular}
\end{center}
\vspace{-3mm}
\caption{Performance comparison of the proposed PIP-compatible framework with SOTA deep CNNs for HSI. B-CNN-3D and C-CNN-3D denote the baseline and custom CNN-3D models respectively (same notations for CNN-32H).}
\label{tab:hsi_comparison}
\vspace{-3mm}
\end{table}

\vspace{-3mm}
\section{Experimental Results}
\label{sec:results}

\subsection{Experimental Setup}

\textit{Model Architectures}: We evaluated the efficacy of our PIP approach on two SOTA HSI models. They were proposed in \cite{datta2021hsi} and consist of a 6-layered 3D CNN, and a hybrid fusion of 3D (one layer) and 2D (two layers) convolutional layers respectively, with a linear classifier layer at the end. The latter has a global average pooling layer before the classifier layer to downsample the last convolutional activation map. We refer to these architectures as our baseline CNN-3D and CNN-32H and they use a patch size of $5{\times}5$ and $3{\times}3$, respectively \cite{datta2021hsi}.

\textit{Datasets}: Our primary benchmark to evaluate our proposal is the HyRANK dataset 
\cite{zhe2021hsi} which is currently the most challenging open-source HSI dataset. It consists of two scenes, namely `Dioni' that is used for training and `Loukia' that is used for testing.  We also used two other publicly available datasets, namely Indian Pines, 
and Salinas Scene \cite{zhe2021hsi,datta2021hsi}, 
where we randomly sample 50\% of the images for training and use the rest for testing. 

\textit{Hyperparameters and Metrices}: We performed full-precision training of our baseline and custom models for $100$ epochs using the standard SGD optimizer with momentum of $0.9$ and an initial learning rate of $0.01$ that decayed by a factor of $10$ after $60$, $80$, and $90$ epochs. We report the overall accuracy (OA), average accuracy (AA), and Kappa coefficient measures to evaluate the HSI performance of our proposed models, simiar to \cite{hamida2018-3d}.

\begin{figure*}
\centering
\includegraphics[width = 1.0\linewidth]{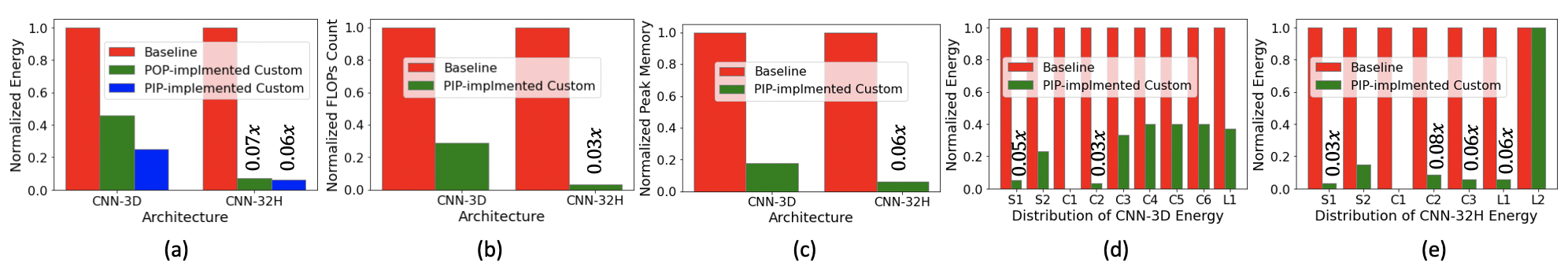}
\vspace{-8mm}
\caption{Comparison of (a) energy consumption, (b) FLOPs count, and (c) peak memory usage between our CNN-3D and CNN-32H-based baseline and custom models for the HyRANK dataset. The distributions of the total energy for the CNN-3D and CNN-32H architecture are shown in (d) and (e) respectively. We denote S1 and S2 as the sensing and sensor-to-SoC communication energy respectively; C1-C6 denote the energies consumed by the convolutional layers, and L1 and L2 denote the classifier layer energies.}
\label{fig:energy_analysis}
\vspace{-2mm}
\end{figure*}

\vspace{-3mm}
\subsection{Quantification of Compression}

We tune the training hyperparameters $C_l^o$, $N$, and $s$ in each dimension such that our custom versions of the CNN-3D and CNN-32H models are within $0.5\%$ of their baseline counterparts and the compression $C$ is maximized. Note that $C$ is computed by plugging in their values shown in Table \ref{tab:hsi_compression}, along with $k$ and $p$, in Eqs. \ref{eq:pip_compression} and  \ref{eq:input_output_mapping}. We choose $k = 3$ for all the three spatial dimensions. A larger $k$ increases the energy consumption of a PIP-implemented convolution, while smaller $k$ reduces the representation capacity and thus model accuracy.
We chose $p = 0$ because increasing $p$ expands the output dimensions and decreases $C$. Table \ref{tab:hsi_compression} shows that $C$ ranges from $5{-}10\times$ over our model architectures and datasets. 

\vspace{-4mm}
\subsection{Classification accuracy}
The accuracies obtained by our baseline and custom models (that captures the specific behavior of the PIP circuits) for the three HSI benchmarks are compared with the other SOTA deep CNNs in Table \ref{tab:hsi_comparison}. Our custom models yield test AAs within $0.54\%$ and $0.83\%$ of the baseline and SOTA models respectively over the three benchmarks. We also ablate the impact of our various compression techniques on the model AAs. For the CNN-3D architecture over Indian Pines dataset, reducing the number of output channels by $10\times$ ($20$ in baseline to $2$ in custom model as shown in Table \ref{tab:hsi_compression}) degrades the AA by $0.1\%$. Increasing the stride in the spectral dimension from $1$ in baseline to $3$ improves the AA by $0.12\%$. This increase is probably due to the effect of regularization. Quantizing the ReLU outputs of the first layer inside the pixel array to $6$-bits via QAT does not cause any further AA drop. Lastly, replacing the element-wise multiplications in the first convolutional layer of the quantized, strided, and low-channel model, with our custom function reduces the AA by $0.50\%$.   


\vspace{-1mm}
\subsection{Analysis of energy-efficiency}

We calculate the total number of floating point operations (FLOPs count), peak memory usage, and the energy consumption of our baseline (processed completely outside the pixel array) and PIP-implemented custom models for the HyRANK dataset. The FLOPs count is computed as the total number of multiply-and-accumulate (MAC) operations in the convolutional and linear layers, similar to \cite{datta2021training,datta2021deep,Kundu_2021_WACV}, while the peak memory is evaluated using the same convention as \cite{chowdhery2019visual}. The total energy is computed as the sum of the image sensor energy, the sensor-to-SoC communication energy obtained from \cite{kodukula2020sensors}, and the energy incurred in processing the CNN layers. Note that the sensor energy is the sum of the pixel array energy\footnote{The pixel array energy is equal to the image read-out energy for the baseline models and in-pixel convolution energy for custom models.} that is obtained from our circuit simulations and the ADC energy that is obtained from \cite{gonugondla2021imc}. We compute the energy for a 3D ($E^{3D}_l$) and 2D ($E^{2D}_l$) convolutional layer $l$  as 
\begin{align}\label{eq:energy_hsi}
     E^{3D}_{l} &= C_{l}^iC_{l}^ok_l^3E_{read}+C_{l}^iC_{l}^ok^3H_l^oW_l^oD_l^oE_{mac}  \\
     E^{2D}_{l} &= C_{l}^iC_{l}^ok_l^2E_{read}+C_{l}^iC_{l}^ok^2H_l^oW_l^oE_{mac}
     \end{align}
\noindent
where $E_{read}$ denotes the energy incurred in reading each element from the on-chip memory to the processing unit and $E_{mac}$ denotes the energy consumed in each MAC operation. Their values are obtained from \cite{kang2018imvlsi}, applying voltage scaling for iso-V$_{dd}$ conditions with other energy estimations\footnote{The energy model for 2D convolutional layers can be extended to linear layers with $k=H_l^o=W_l^o=1$ and $C_l^i$ and $C_l^o$ as the number of input and output neurons respectively.}. Other notations in Eq. \ref{eq:energy_hsi} are obtained by appending a subscript $l$ to the notations used in Section \ref{subsec:pip_compression} to reflect the $l^{th}$ layer.

As shown in Fig. \ref{fig:energy_analysis}(a)-(c), our CNN-3D-based compressed custom model yields $4.0\times$, $3.4\times$, and $5.5\times$ reduction in the FLOPs count, peak memory, and energy respectively compared to the baseline counterpart. For CNN-32H, the reduction factors are $16.7\times$, $30.3\times$, and $16.6\times$.

In order to quantify the energy benefits of PIP alone, we also estimate the energy of our custom models, \underline{p}rocessed completely \underline{o}utside the \underline{p}ixel (POP) array. Fig. \ref{fig:energy_analysis} shows that the PIP-implemented custom models lead to a $1.5\times$ reduction in the total energy on average across the two architectures, compared to its POP-implemented counterparts. We also show the component-wise energy reductions in our PIP-implemented compressed model for CNN-3D and CNN-32H in Fig. \ref{fig:energy_analysis}(d)-(e). As we can see, our $10\times$ compression in CNN-3D reduces the total energy incurred in CNN processing by $3.63\times$, while the PIP implementation contributes to a reduction of $19.23\times$ in the sensing energy. On the other hand, the reduction factors in CNN-32H are estimated to be $4.17\times$ and $31.25\times$, respectively.

\vspace{-4mm}
\section{Summary \& Conclusions}\label{sec:conc}

In this paper, we propose a PIP solution that can reduce the bandwidth and energy bottleneck associated with the large amount of data transfer in HSI pipelines. We present two PIP-optimized CNN models that are obtained via our algorithm-hardware co-design framework. This framework captures the non-idealities associated with our PIP implementation and compresses the HSIs significantly within the first few layers. Our approach reduces the energy incurred in the image sensor, the data transfer energy between sensing and processing, and the compute energy of the downstream processing. 
Our PIP-enabled custom models are estimated to reduce the total average energy consumed for processing the standard HSI benchmarks by $10.3\times$ compared to SOTA in-sensor processing solutions, while yielding average accuracies within $0.56\%$ of the baseline models. 

\bibliographystyle{IEEEbib}
\bibliography{strings,refs}

\end{document}